\pdfoutput=1
\documentclass[
 reprint,
superscriptaddress,
groupedaddress,
showpacs,preprintnumbers,showkeys,
 amsmath,amssymb,
 aps,
pra,
]{revtex4-1}

\usepackage{graphicx,stmaryrd}
\usepackage{dcolumn}
\usepackage{bbold}
\usepackage{bm}
\usepackage{color}
\usepackage{./diagram}
\usepackage{hyperref}
\newcommand{\n}[1]{\begin{normalsize}#1\end{normalsize}} 
\newcommand{\s}[1]{\begin{small}#1\end{small}} 
\newcommand{\f}[1]{\begin{footnotesize}#1\end{footnotesize}} 
\definecolor{linkblue}{rgb}{0.1,0.3,.7}
\definecolor{forestgreen(web)}{rgb}{0.13, 0.55, 0.13}
\definecolor{lava}{rgb}{0.81, 0.06, 0.13}
\hypersetup{
breaklinks,
    colorlinks,
    citecolor=forestgreen(web),
    filecolor=linkblue,
    linkcolor=lava,
    urlcolor=linkblue
}

\begin{document}


\title{Loops in Anti--de Sitter Space}

\author{Igor Bertan}
\email{igor.bertan@physik.lmu.de}
\author{Ivo Sachs}%
 \email{ivo.sachs@physik.lmu.de}
\affiliation{\vspace{3pt}
 Arnold Sommerfeld Center for Theoretical Physics,\\ Ludwig Maximilian University of Munich,\\
Theresienstra{\ss}e\enspace 37, D-80333 M\"unchen,\enspace Germany
  \vspace{6pt}
}%



\date{\today}

\begin{abstract}
We obtain analytic results for the four-point amplitude, at one loop, of an interacting scalar field theory in four-dimensional, Euclidean anti--de Sitter space without exerting any conformal field theory knowledge. For the two-point function, we provide analytic expressions up to two loops. In addition, we argue that the critical exponents of correlation functions near the conformal boundary of anti--de Sitter space provide the necessary data for the renormalization conditions, thus replacing the usual on-shell condition.\\

\noindent \f{DOI: \href{https://doi.org/10.1103/PhysRevLett.121.101601}{10.1103/PhysRevLett.121.101601}}
\end{abstract}

\maketitle

\section{\label{sec:level1}Introduction}
Over the past sixty years, there has been tremendous progress in the calculation of scattering amplitudes in quantum field theory---in particular, concerning higher loop amplitudes in Yang-Mills theory and (super)gravity. 

At the same time, we have very few analytic results for loop amplitudes in curved space-times. More precisely, while their short-distance expansion and, in particular, the structure of counterterms  they give rise to is rather well known \cite{Birrell:1982ix,Comellas:1994da}, we know little about their dependence on coordinates beyond that (e.g., Ref. \cite{Akhmedov:2013vka} and references therein). Even in de Sitter (dS) or anti--de Sitter (AdS) space, which are maximally symmetric, admitting the same number of isometries as Minkowski space, little is known about such amplitudes; see Refs. \cite{Penedones:2010ue,Fitzpatrick:2011hu,Aharony:2016dwx,Rastelli:2016nze,Cardona:2017tsw,Giombi:2017hpr,Yuan:2018qva} for recent progress. The reason for this is that, while in Minkowski space, the momentum representation leads to a hierarchy of elementary integrals; in dS or AdS, this is not the case, and the coordinate representation generally leads to integral expressions that are more manageable in (A)dS. Still, except for some special cases, we lack the technical tools for performing the integrations completely. 

In this Letter we report on some progress considering the simplest interacting renormalizable field theory. Concretely, we compute the two- and four-point functions for $\lambda\phi^4$ theory \footnote{\s{The cubic coupling is known to be extremal, and the simplest $\Yup$-diagram is divergent}} to the second order in the coupling $\lambda$ on the Poincar\'e patch of Euclidean AdS$_4$ by explicitly evaluating the corresponding one- and two-loop integrals in coordinate representation. Working on AdS, we avoid complications that arise from IR effects on dS, for instance \cite{Akhmedov:2013vka}. It turns out that even this simplified setting is beyond reach for external legs at generic points in AdS,\newpage \noindent but for insertions on the conformal boundary we are able to get explicit expressions. 

Being able to go beyond the short-distance expansion, we encounter an interesting complication concerning the renormalization conditions: For distances that are small compared to the curvature scale, the problem reduces to that in flat space, and the physical masses provide the right boundary conditions for the renormalized propagator, for instance. At scales of the order of the curvature radius, however, there is no meaningful  definition of the mass of a scalar field, and one needs to identify a reasonable renormalization condition. In the present case, we will find that the critical exponents of correlation functions at the conformal boundary of AdS provide just that. Indeed, the bulk amplitudes on AdS with external legs inserted at the boundary define a crossing symmetric point correlation function of some hitherto unknown primary operator of a conformal field theory (CFT) on the conformal boundary by construction, and therefore a consistent CFT. Of course, we do not know what is the microscopic realization of this CFT, nor do we need it. What matters is that a primary operator has a well-defined dimension which is given by the critical exponent of its correlation function near the conformal boundary. This is what replaces the physical mass at large distances (see also Ref. \cite{Beisert:2010jr}). 

Concretely, let us consider a scalar field with classical action \footnote{\s{$\mathrm{d}\mu_{x_1,x_2,\ldots}=\mathrm{d}^4 x_1 \sqrt{|g(x_1)|}\times \mathrm{d}^4x_2 \sqrt{|g(x_2)|}\dots$}}
\n{
\begin{equation}
    S=- \int \mathrm{d}\mu_x \left( \frac12 (\partial\phi)^2 +\frac{m^2}{2}  \phi^2+\frac{\lambda}{4!} \phi^4\right)
\end{equation}}%
on the Poincar\'e patch of hyperbolic space of radius $1/a$ with the metric
\n{
\begin{equation}
\label{poincmetricc}
\mathrm{d}s^2 =\frac{1}{a^2 z^2} (\mathrm{d}z^2 + {\mathrm{d}x^{i}}^2),~~~ (z,x^i) \in (\mathbb{R}_{\geq 0}, \mathbb{R}^3).
\end{equation}}

There are two admissible boundary conditions for the classical scalar field corresponding to the asymptotic behavior $\phi(z,x^i)\sim z^\Delta\varphi(x^i)$ with $a^2 \Delta(\Delta-3)=m^2$. Here we will focus on the conformally coupled scalar for which $m^2 = - 2a^2$ and therefore $\Delta=1,2$. For $\Delta=1$, there are further complications due to infrared divergencies. We thus focus on $\Delta=2$ in this case. The scalar propagator is then given by (cf. \cite{Birrell:1982ix} and references therein)
\n{\begin{equation}
\label{prop}
\Lambda(K) =\frac{a^2 K^2}{4\pi^2 (1-K^2)}\,,
\end{equation} }
where $K$ is the invariant bilocal function 
\begin{equation}
\label{uinpoinc}
K \equiv K_{x,y}= \frac{2zw}{(x^i - y^i)^2 + z^2 + w^2}
\end{equation}
with coordinates $x^\mu=(z,x^i)$ and $y^\mu = \left( w, y^i\right)$. Henceforth we denote $x^2 \equiv {x^i}^2$. Taking one point to the conformal boundary, $z\sim 0$, then 
\begin{equation}
\label{boundbeh}
 \frac{K}{z} 
  \sim \bar{K} = \frac{2w}{(x - y)^2 + w^2}
 \end{equation}
reduces to the usual bulk-to-boundary propagator \cite{Witten:1998qj}. 
\subsection{\label{twopoint}Two-Point Function}
To order $\lambda^2$, the two-point function contains the following fundamental constituents:
\begin{align}
\label{1PI2pt}
\begin{split}
\diagramone
\end{split}
\end{align}
where the first diagram is just a mass shift which will, however, play a prominent role in the following. It corresponds to the elementary integral (here $K=K_{x_1,x_2}$)
\begin{align}
      \label{massshiftsol}
    \begin{split}
\mathcal{I}_2 =& \int \mathrm{d}\mu_x \Lambda(x_1,x) \Lambda(x_2,x)\\
=& - \frac{K}{8\pi^2 (1-K^2)} \left(  \log \frac{1-K}{1+K} + K\log\frac{4 K^2}{1-K^2}\right)\!.
    \end{split}
\end{align}
\subsubsection{Tadpole Diagrams}
The one-loop tadpole diagram  
\begin{equation}
\label{tadpolediag}
\mathcal{H}_2 = \int \mathrm{d}\mu_x  \Lambda(x,x_1) \Lambda(x,x_2)\Lambda(x,x)
\end{equation}
requires regularization at short distances, $K\rightarrow 1$. We choose
\begin{equation}
\label{KtoKepsilon}
    K \rightarrow \frac{K}{1+\epsilon}\,,
\end{equation}
which cuts out a small $\epsilon$-ball around the pole in the propagator and rescales it by $1/(1+\epsilon)$. With this, the  tadpole diagram reduces to a mass counterterm as expected:
\begin{equation}
\label{curlHcomputed}
\mathcal{H}_2 = \frac{a^2 }{8\pi^2}\left(\frac{1}{\epsilon} - \frac{9}{2} \right)\mathcal{I}_2 + \mathcal{O}(\epsilon).
\end{equation}
For the two-loop tadpole diagram 
\begin{equation}
\label{dtdint}
    \mathcal{L}_2 = \int \mathrm{d}\mu_{x,y}  \Lambda(x_1,x) \Lambda(x_2,x) \Lambda(x,y)^2 \Lambda(y,y),
\end{equation}
we adopt the same regularization as above. It is then possible to show that for $z\neq 0$, the $y$ integral is well defined and is independent of $x$. Consequently, the nested integral  (\ref{dtdint}) factorizes as 
\begin{equation}
    \mathcal{L}_2 = \frac{ a^2 }{(4\pi^2)^3}\frac{1}{\epsilon (2+\epsilon)(1+\epsilon)^{4}}~\mathcal{M}_2\times \mathcal{I}_2,
\end{equation}
where $\mathcal{I}_2$ is again the mass shift. Using translation invariance to set $x=(1,0)$, we find
\begin{equation}
    \mathcal{M}_2 = 8  \int_{-\infty}^{\infty} \mathrm{d}^4 y~ \frac{[y^2 +(w +1)^2 + \epsilon Q ]^{-2}}{[y^2 +(w -1)^2 + \epsilon Q ]^2}\,,\nonumber
\end{equation}
with $Q= y^2 + w^2 +1$. This integral is most easily computed using Schwinger parameters. 
For small $\epsilon$, we then end up with 
\begin{equation}\notag
    \mathcal{L}_2 = \frac{ a^2 \pi^2 }{2 (4\pi^2)^3} \left( \frac{14+13 \log \frac{\epsilon}{2}}{2}-\frac{1+\log\frac{\epsilon}{2}}{\epsilon } \right) \mathcal{I}_2+\mathcal{O}(\epsilon).
\end{equation}
\subsubsection{Sunset Diagram}
\label{sunsetsection}
Finally, we consider the sunset diagram
\begin{equation}
\label{K2integral}
\mathcal{K}_2 =    \int \mathrm{d} \mu_{x,y}\Lambda(x_1,x) \Lambda(x,y)^3  \Lambda(x_2,y).
\end{equation}
Let us first consider the subdiagram with only one external leg attached:
\begin{equation}
\label{jellyfishbeginning}
\mathcal{J}_2 = \int \mathrm{d}\mu_y \Lambda (x,y)^3 \Lambda (x_2,y).
\end{equation}
If we denote the restriction of $\mathcal{J}_2$ to the conformal boundary by $J_2$, we have
\begin{equation}
J_2 = \frac{2^7 a^4 z_2^2 z''^6}{(4\pi^2)^4(1+\epsilon)^4}\int_{-\infty}^{\infty} \mathrm{d}^4 y  ~ \frac{w^4 [Q+ 2 z'' w + \epsilon Q]^{-3}}{[Q - 2 z'' w+\epsilon Q]^3}\, ,\nonumber
\end{equation}
where $Q = y^2 + z''^2 + w^2$. Here we use translation invariance, as above, to set $x_2'=(0, 0)$, followed by an inversion  with $z'' = z'/(x'^2 + z'^2)$ \cite{Freedman:1998tz}.
This integral is again evaluated using Schwinger, leading to
\begin{equation}
J_2 = \frac{a^4  \pi^2}{4 (4 \pi^2)^4}   K^2_{x, x_2}\left(\frac{1}{\epsilon} +3  \log \frac{\epsilon}{2} -\frac{1}{2} \right)\!,
\end{equation}
where we use the fact that $2 z''= \bar{K}_{x, x_2}$ to recover the covariant form. 
The full sunset diagram can now be obtained by attaching the remaining leg to $J_2$.  This yields
\begin{equation}
\label{sunsetfull}
\mathcal{K}_2  = \frac{a^2  \pi^2}{4(4\pi^2)^3} \left(\frac{1}{\epsilon} +3  \log \frac{\epsilon}{2} -\frac12 \right) \mathcal{I}_2,
\end{equation}
where $\mathcal{I}_2$ is the mass shift (\ref{massshiftsol}), and where we use the symmetry of $\mathcal{K}_2$ under permutation of the external legs to construct the unique extension of the above correlator in the bulk. 

In conclusion, all diagrams that contribute to the two-point function, up to the second order in the coupling constant, reduce to the mass shift diagram which relates the Lagrangian mass to the conformal dimension of a primary operator, $\mathcal{O}$, when evaluated at the boundary. 
This suggests replacing the renormalization condition defining the physical mass, which itself is not well defined in AdS, with a renormalization condition on the conformal dimension of $\mathcal{O}$. Our choice sets $\Delta$ to $2$ at all orders in the perturbation. Below, we will see that this is consistent with the four-point function.

\subsection{Four-Point Function}
Up to second order in $\lambda$, the one-particle irreducible diagrams contributing to the four-point function are
\begin{equation}
    \diagramtwo
\end{equation}
The tree-level contribution of the quartic vertex to the four-point function, given by 
\begin{equation}
\mathcal{I}_4 = \int \mathrm{d}\mu_x  \Lambda(x_1,x)\Lambda(x_2,x)\Lambda(x_3,x)\Lambda(x_4,x),
\end{equation}
has already been calculated for external legs inserted on the boundary. Here we just quote the result \cite{Muck:1998rr}:
\begin{align}
\label{finalformula}
I_4& = \frac{4^2 a^4 (\Pi_{i=1}^4 z_i)^2 }{(4\pi^2)^3 (\eta \zeta \Pi_{i<j} r_{ij})^{\frac{4}{3}}} \\
&\times \int_{0}^{\infty} \mathrm{d}z \,{}_2F_1\!\left[2,2;4; 1 - \left(\frac{\eta  + \zeta }{\eta  \zeta }\right)^2 - \frac{4\sinh^2z}{\eta  \zeta}\right]\nonumber\!,
\end{align}
where we introduce $r_{ij}=|x_i-x_j|$ and the conformal cross ratios of the coordinates on the boundary $\eta = r_{14}r_{23}/r_{12}r_{34},~ \zeta = r_{14}r_{23}/r_{13}r_{24}$.

\subsubsection{Loop Diagram}
Let us calculate the one-loop correction given by the double integral
\begin{equation}\notag
\mathcal{K}_4 = \int \mathrm{d}\mu_{x,y} \Lambda(x_1,x)\Lambda(x_2,x)\Lambda(x,y)^2\Lambda(x_3,y)\Lambda(x_4,y),
\end{equation}
considering again first the simpler integral
\begin{equation}
\label{simplerint}
\mathcal{J}_4 = \int \mathrm{d} \mu_y \Lambda (x,y)^2 \Lambda (x_3,y)\Lambda (x_4,y).
\end{equation}
By sending $x_3$ and $x_4$ to the boundary, it takes the form
\begin{equation}
J_4= \frac{a^8 ( z_3 z_4)^2}{(4\pi^2)^4(1+\epsilon)^{8}} \int \mathrm{d}\mu_y \frac{\bar{K}_{x_3, y}^2 \bar{K}_{x_4, y}^2 (1+K_{x,y}+\epsilon)^{-2}}{K_{x,y}^{-4}(1-K_{x,y}+\epsilon)^2 }\notag\,.
\end{equation}
We may safely set $\epsilon=0$ in the prefactor, since the integral diverges logarithmically. As before, we translate the points $x, x_3, x_4$ by $(0,- x^i_4)$, which gives $x_4'=(0, 0)$, and we perform an inversion with the inverted points denoted  by double primes so that
\begin{equation}
\label{J4nu}
J_4 = \frac{a^4 z''^{4} (z_3 z_4)^2}{2^{-7}(4\pi^2)^4 r_{34}^{4}} \int_{-\infty}^{\infty} \frac{\mathrm{d}^4 y}{w^{-4}}~ \frac{[(Q_{-} + \epsilon Q) (Q_{+}+\epsilon Q)]^{-2}}{[(x_3'' - x''-y)^2 + w^2]^2}\,,\notag
\end{equation}
where $Q_{\pm} = y^2 + (z'' \pm w)^2 $ and $Q=y^2 + z''^2 + w^2$. Integrating yields 
\begin{equation}\label{nl1}
J_4 = \frac{a^4  \pi^2}{(4 \pi^2)^4}   K^2_{x, x_3} K^2_{x, x_4}\left(\log\frac{\alpha^2+1}{ 2\epsilon} -2 \right)\!,
\end{equation}
where
\begin{equation}
 \frac{4}{\alpha^2 + 1}\equiv r_{34}^2 \bar{K}_{x, x_3} \bar{K}_{x, x_4}=\frac{4 z''^2}{(x_3'' - x'')^2 + z''^2}\,.
\end{equation} 

Finally, we attach the remaining two external legs to (\ref{simplerint}). Sending all $x_i$ to the boundary, we have
\begin{align}
\begin{split}
K_4 = \frac{4^4 a^4  ( \Pi_{i=1}^4 z_i)^2 \pi^2}{2 (4 \pi^2)^6}  \int_{-\infty}^{\infty} \mathrm{d}^4x ~  \frac{ z^{4} \left(\log\frac{\alpha^2+1}{2\epsilon} -2 \right)}{\Pi_{i=1}^4[(x_i-x)^2 + z^2]^2 }\notag\,.
\end{split}
\end{align}
We then repeat the by-now-familiar procedure by translating $x_k,~k = 1,\dots,4$ by $(0,- x^i_4)$ (denoted by primes), inverting all points (denoted by double primes), and then making the substitution $(z'',{x''}^{i})= (z, x^i + {x''_3}^{i})$. The integration is then performed in the usual fashion, leading to 
\begin{equation}
\label{K4final}
K_4 = \frac{1}{16\pi^2} \left[-I_4 \left(\frac{11}{3}+ \log \frac{\epsilon}{8}  \right) +  L_4 \right]\!,
\end{equation}
where
\begin{align*}
\label{Lfour}
\begin{split}
    L_4 =& \frac{3\times 4^2 a^4 (\Pi_{i=1}^4 z_i)^2}{2(4\pi^2)^3(\eta\zeta ~\Pi_{i<j} r_{ij})^{\frac{4}{3}}}\\
    \times &\int_{0}^{\infty} \mathrm{d}s \int_0^1 \mathrm{d}r  \frac{ [s r (1-r)]\log (1+s)}{ (1+s)^{2} [\frac{s r (1-r) }{ \eta^2} +\frac{r}{\zeta^2} +1- r]^{2}}\,.
    \end{split}
\end{align*}
This is the main result of this Letter. To continue, $L_4$ can be evaluated numerically or, alternatively, order by order in an expansion in $1-\zeta^{-2}$ and $\eta^{-1}$. Then the expansion coefficients of (\ref{K4final}) contain important physical information, which can be extracted by comparing  them with the operator product expansion (OPE) in CFT. 
\subsection{Comparison to Conformal Field Theory}
In flat space, loop corrections to the tree-level amplitudes contain information about the coupling dependence of the masses of resonances, for instance. In AdS, where there is no scattering, the role of physical masses is taken by the dimensions of operators of some CFT dual  \cite{Maldacena:1997re,Gubser:1998bc,Witten:1998qj}. For a scalar field $\phi$ in AdS, this CFT is characterized by the existence of a scalar operator $\mathcal{O}$ dual to $\phi$. Our renormalization scheme fixes its two-point function to 
\begin{equation}
    \langle \mathcal{O}(x_1) \mathcal{O}(x_2) \rangle =\diagrama =  \frac{N_\phi}{r_{12}^{4}}\,,
\end{equation}
where $N_\phi = a^2 (2 z_1 z_2)^2/4\pi^2$. To continue, we take $z_i=z\sim 0$ for all external legs. 
Then, expanding the holographic four-point function in the variables $\eta^{-1}$ and $Y= 1- \zeta^{-2}$ yields
\begin{align}
\label{aaaa1}
    \begin{split}
    &\langle \mathcal{O}(x_1)  \mathcal{O}(x_2) \mathcal{O}(x_3) \mathcal{O}(x_4) \rangle =\\
    =&~3 \times \diagramb +\lambda~ \diagramc + \mathcal{O}\left( \lambda^3 \right)\\
    = &\frac{N_\phi^2}{(r_{12}r_{34})^{4}} \left[1 + \frac{1}{\eta^4} \sum_{l,m=0}^\infty F_{lm}(\log\eta,\lambda_R)~ \frac{Y^m}{\eta^{2 l}}\right]\!,
        \end{split}
\end{align}
where each $F_{lm}$ can be derived from the results of the previous sections. Note that the factor ``3" in the diagrams merely indicates that there are three diagrams of this type that contribute to the correlator. Here we furthermore  introduce the renormalized coupling constant by a nonminimal subtraction of the form
\begin{equation}
    \lambda = \lambda_R + \frac{ \lambda_R^2}{32 \pi^2 } \left( 5 + 3 \log\frac{\epsilon}{8} \right)+ \mathcal{O}(\lambda_R^3).
\end{equation}
At this point, we should emphasize that a nonvanishing beta function of the bulk theory does not spoil the conformal symmetry on the boundary, since scale transformations on the boundary correspond to translations in AdS.

Let us now consider a general four-point function of identical scalar operators $\mathcal{O}$ of conformal dimension $\nu$ in a three-dimensional CFT. Its decomposition in conformal blocks (CBs) $G_s^{\nu_s}$ is \cite{Dolan:2000ut} 
\begin{equation}
\label{CBD}
\langle \mathcal{O}(x_1) \mathcal{O}(x_2) \mathcal{O}(x_3)\mathcal{O}(x_4) \rangle = \sum_{s,\nu_s}\frac{ C_s^{\nu_s}~ G_s^{\nu_s}(\eta,Y)}{(r_{12}r_{34})^4 ~\eta^{\nu_s - s}}\, ,
\end{equation}
where the only unknowns are the spectrum of the CFT--- i.e., the spin $s$ and conformal dimension $\nu_s$ of all primary operators, as well as the associated OPE coefficients $C_s^{\nu_s}$. These can, in turn, be determined by comparison with (\ref{aaaa1}). Our recipe is the following: we first compute $F_{lm}$ and identify the set of possibly contributing CBs at that order in $\eta$ and $Y$. Afterwards, we expand (\ref{CBD}) in the conformal dimensions $\nu_s =\bar{\nu}_s +  \lambda_R \gamma_s^{\bar{\nu}_s(1)}+ \lambda_R^2 \gamma_s^{\bar{\nu}_s(2)} + \mathcal{O}(\lambda^3_R)$ and OPE coefficients $C_s^{\nu_s} =C^{\bar{\nu}_s (0)}_{s} +  \lambda_R C^{\bar{\nu}_s (1)}_{s}+ \lambda_R^2 C^{\bar{\nu}_s (2)}_{s} + \mathcal{O}(\lambda^3_R)$. This allows for a direct comparison with $F_{lm}$ at different orders in $\lambda_R$.

For brevity, in this Letter we will focus on the terms with $(l,m) = {(0,0),(1,0),(0,1),(0,2)}$. The CBs which contribute at these orders are
\begin{align*}
    G_0^{\nu_0} &=1 +\frac{\nu_0}{4}Y+ \frac{\nu_0^3(\nu_0 +1)^{-1}}{8(2\nu_0-1)\eta^2} + \frac{\nu_0(\nu_0+2)^2}{32(\nu_0+1)}Y^2,\\
    G_1^{\nu_1} &=- \frac{1}{2}Y - \frac{\nu_1+1}{8} Y^2, \\
    G_2^{\nu_2} &= - \frac{1}{2\eta^2} + \frac{3}{8} Y^2.
\end{align*}
Note that the identity operator $\mathbb{1}$ appears in (\ref{aaaa1}) in the form of the $s$-channel disconnected diagram, as its CB reads $G_0^0=1$. Thus, $C_0^0 = N_\phi^2$.

For $(l,m) =(0,0)$, one finds
\begin{align*}
    F_{00} =& ~2 +\frac{\lambda_R}{48 \pi^2} \left(-1 +6\log \eta \right)\\
    +& \frac{\lambda_R^2}{3 \times 2^8 \pi^4}   \left(5 - 11 \log \eta + 3(\log \eta)^2  \right)\!.
\end{align*}
The only contribution comes from the primary operator $:\mkern-4mu\mathcal{O}^2\mkern-4mu:$ having conformal dimension $\bar{\nu}_{0}=4$. Then, comparing the corresponding CB expansion with $F_{00}$ at lowest order in the coupling constant yields $C^{4(0)}_{0} = 2 N_\phi^2$.
At first order in $\lambda_R$, we get
\begin{equation*}
\gamma_0^{4(1)} = -  \frac{1}{16 \pi^2}\,,\qquad C^{4(1)}_{0} = -  \frac{N_\phi^2}{48 \pi^2}\,;
\end{equation*}
whereas at second order in $\lambda_R$, we find
\begin{align*}
\gamma_0^{4(1)} =& \pm \frac{1}{16 \pi^2}\,,\qquad~ \gamma_0^{4(2)} = \frac{ 5 }{3 \times 2^8 \pi^4}\,,\\
C_{0}^{4(2)} =& \frac{5 N_\phi^2}{3\times 2^8 \pi^4}\,.
\end{align*}
Note that $\gamma_0^{4(1)} $ agrees at different orders in $\lambda_R$. This provides an important consitency test for the AdS/CFT duality beyond tree level in the bulk. In previous work, this property was taken as part of the definition of loop diagrams in the bulk (e.g., Refs. \cite{Aharony:2016dwx,Fitzpatrick:2011hu,Rastelli:2016nze}). It is reassuring to see that it is indeed compatible with an actual bulk calculation. Put differently, this supports the argument that CFT does indeed describe the structure underlying amplitudes of QFT in AdS rather than acting merely as a definition of some bulk theory specified by its correlation functions.

For $(l,m) = (0,1)$, $F_{01}$ reads
\begin{align*}
    F_{01} =& ~2 +\frac{\lambda_R}{96 \pi^2} \left(-5 + 12\log \eta \right)\\
    +& \frac{\lambda_R^2}{12 \times 2^8 \pi^4}   \left(31 - 50 \log \eta + 12(\log \eta)^2  \right)\!.
\end{align*}
In addition to $:\mkern-9mu\mathcal{O}^2\mkern-9mu:\,$, there might be a contribution of a vector operator of dimension $\bar{\nu}_1 = 5$. However, by comparing $F_{01}$ with the expansions of the CBs, one can infer that the vector operator does not appear in the OPE. This agrees with our expectation based on general CFT arguments.

The term satisfying $(l,m)=(0,2)$ reads
\begin{align*}
    F_{02} =& ~3+ \frac{9 \lambda_R}{80 \pi^2} \left(-\frac{11}{20} + \log \eta \right)\\
    +&  \frac{\lambda_R^2}{500 \times 2^8  \pi^4 }   \left(1408 - 1965 \log \eta + 450 (\log \eta)^2 \right)\!.
\end{align*}
Bearing in mind that the vector operator of dimension $5$ does not appear, the only new operator which contributes is the spin-$2$ primary of the schematic form $:\mkern-6mu\mathcal{O}\partial^i \partial^j \mathcal{O}\mkern-6mu:$ with $\bar{\nu}_{2}=6$. It follows that
\begin{align*}
C_{2}^{6(0)} &= \frac{16 N_\phi^2}{5}\,,\qquad \gamma_{2}^{6(1)} =C_{2}^{6(1)} = 0, \\
\gamma_{2}^{6(2)} &= - \frac{1}{20\times 2^8 \pi^4}\,,\qquad C_{2}^{6(2)} = - \frac{11 N_\phi^2}{375 \times 2^8 \pi^4}\,.
\end{align*}
Also, here $\gamma_{2}^{6(1)}$ agrees at both orders in $\lambda_R$. An interesting observation is that the spin-$2$ primary, in spite of not being conserved, does not acquire an anomalous dimension at first order in the coupling $\lambda_R$. However, it does get modified at second order.

The last term we consider here corresponds to $(l,m)= (1,0)$ with
\begin{align*}
    F_{10} =& ~\frac{\lambda_R}{120 \pi^2 } \left(\frac{17}{5} + 12 \log \eta  \right)\\
    +&  \frac{\lambda_R^2}{150 \times 2^8 \pi^4 }  \left(-\frac{491}{5} - 584 \log \eta + 120 (\log \eta)^2 \right)\!,
\end{align*}
which requires a new scalar operator $:\mkern-4mu\mathcal{O} \nabla^2 \mathcal{O}\mkern-4mu:$ of conformal dimension $\bar{\nu}_0=6$ and
\begin{align*}
C_{0}^{6(0)} &= \frac{8 N_\phi^2}{7}\,,~\gamma_{0}^{6(1)} = -\frac{1}{16\pi^2}\,,~C_{0}^{6(1)} = \frac{239 N_\phi^2}{14\times 420 \pi^2}\,, \\
\gamma_{0}^{6(2)} &= \frac{23}{15 \times 2^7 \pi^4}\,,\qquad ~ C_{0}^{6(2)} = - \frac{1637 N_\phi^2}{5 \times 2^6 \times 1029}\,.
\end{align*}\\[0.08in]
Again, there is an agreement of $\gamma_0^{6(1)}$ at different orders in $\lambda_R$. Eventually, this gives a complete characterization of all operators of spin $s \leq 2$ entering the OPE. 

\section{Conclusions}

In this Letter, we computed quantum corrections to the two- and four-point correlation functions up to second order in the coupling constant for the simplest scalar field theory in AdS$_4$. The obtained results for the two- and four-point functions are mutually consistent. Furthermore, the holographic four-point function can systematically be expanded in the conformal invariants to reveal the OPE structure of the dual CFT, along with the corrections to both the OPE coefficients and conformal dimensions of primary operators. This was carried out here at low orders, disclosing a mathematically consistent dual CFT. In particular, the absence of the stress tensor and of any conserved current becomes explicit.

To summarize, the conformally coupled bulk $\lambda\phi^4$ theory describes a one-parameter family of dual conformal field theories whose OPE coefficients and dimensions are parametrized by the renormalized coupling $\lambda_R$. Generalization of the result presented here to boundary conditions with $\Delta=1$ is possible, although with some extra complications concerning infrared divergencies for the tadpole and sunset diagrams. For massless spin-$1$ and spin-$\frac{1}{2}$ particles in AdS, the propagator is again given by (\ref{prop}) modulo parallel transport of the polarization vectors. This means that spin-$1$ and spin-$\frac12$ fields lead to similar integrals to those computed here, and hence QED and scalar QED in AdS can be quantized in the same way.

\begin{acknowledgments}
The authors would like to thank Evgeny Skvortsov for very substantial input and advice. This work was supported by the DFG Transregional Collaborative Research Centre TRR 33 and the DFG cluster of excellence ``Origin and Structure of the Universe".

\end{acknowledgments}
\bibliographystyle{apsrev4-1}
\bibliography{bib.bib}

\begin{thebibliography}{19}%
\makeatletter
\providecommand \@ifxundefined [1]{%
 \@ifx{#1\undefined}
}%
\providecommand \@ifnum [1]{%
 \ifnum #1\expandafter \@firstoftwo
 \else \expandafter \@secondoftwo
 \fi
}%
\providecommand \@ifx [1]{%
 \ifx #1\expandafter \@firstoftwo
 \else \expandafter \@secondoftwo
 \fi
}%
\providecommand \natexlab [1]{#1}%
\providecommand \enquote  [1]{``#1''}%
\providecommand \bibnamefont  [1]{#1}%
\providecommand \bibfnamefont [1]{#1}%
\providecommand \citenamefont [1]{#1}%
\providecommand \href@noop [0]{\@secondoftwo}%
\providecommand \href [0]{\begingroup \@sanitize@url \@href}%
\providecommand \@href[1]{\@@startlink{#1}\@@href}%
\providecommand \@@href[1]{\endgroup#1\@@endlink}%
\providecommand \@sanitize@url [0]{\catcode `\\12\catcode `\$12\catcode
  `\&12\catcode `\#12\catcode `\^12\catcode `\_12\catcode `\%12\relax}%
\providecommand \@@startlink[1]{}%
\providecommand \@@endlink[0]{}%
\providecommand \url  [0]{\begingroup\@sanitize@url \@url }%
\providecommand \@url [1]{\endgroup\@href {#1}{\urlprefix }}%
\providecommand \urlprefix  [0]{URL }%
\providecommand \Eprint [0]{\href }%
\providecommand \doibase [0]{http://dx.doi.org/}%
\providecommand \selectlanguage [0]{\@gobble}%
\providecommand \bibinfo  [0]{\@secondoftwo}%
\providecommand \bibfield  [0]{\@secondoftwo}%
\providecommand \translation [1]{[#1]}%
\providecommand \BibitemOpen [0]{}%
\providecommand \bibitemStop [0]{}%
\providecommand \bibitemNoStop [0]{.\EOS\space}%
\providecommand \EOS [0]{\spacefactor3000\relax}%
\providecommand \BibitemShut  [1]{\csname bibitem#1\endcsname}%
\let\auto@bib@innerbib\@empty
\bibitem [{\citenamefont {Birrell}\ and\ \citenamefont
  {Davies}(1984)}]{Birrell:1982ix}%
  \BibitemOpen
  \bibfield  {author} {\bibinfo {author} {\bibfnamefont {N.~D.}\ \bibnamefont
  {Birrell}}\ and\ \bibinfo {author} {\bibfnamefont {P.~C.~W.}\ \bibnamefont
  {Davies}},\ }\href {\doibase 10.1017/CBO9780511622632} {\emph {\bibinfo
  {title} {{Quantum Fields in Curved Space}}}},\ Cambridge Monographs on
  Mathematical Physics\ (\bibinfo  {publisher} {Cambridge Univ. Press},\
  \bibinfo {address} {Cambridge, UK},\ \bibinfo {year} {1984})\BibitemShut
  {NoStop}%
\bibitem [{\citenamefont {Comellas}\ \emph {et~al.}(1995)\citenamefont
  {Comellas}, \citenamefont {Haagensen},\ and\ \citenamefont
  {Latorre}}]{Comellas:1994da}%
  \BibitemOpen
  \bibfield  {author} {\bibinfo {author} {\bibfnamefont {J.}~\bibnamefont
  {Comellas}}, \bibinfo {author} {\bibfnamefont {P.~E.}\ \bibnamefont
  {Haagensen}}, \ and\ \bibinfo {author} {\bibfnamefont {J.~I.}\ \bibnamefont
  {Latorre}},\ }\href {\doibase 10.1142/S0217751X95001339} {\bibfield
  {journal} {\bibinfo  {journal} {Int. J. Mod. Phys.}\ }\textbf {\bibinfo
  {volume} {A10}},\ \bibinfo {pages} {2819} (\bibinfo {year} {1995})},\ \Eprint
  {http://arxiv.org/abs/hep-th/9404080} {arXiv:hep-th/9404080} \BibitemShut
  {NoStop}%
\bibitem [{\citenamefont {Akhmedov}(2014)}]{Akhmedov:2013vka}%
  \BibitemOpen
  \bibfield  {author} {\bibinfo {author} {\bibfnamefont {E.~T.}\ \bibnamefont
  {Akhmedov}},\ }\href {\doibase 10.1142/S0218271814300018} {\bibfield
  {journal} {\bibinfo  {journal} {Int. J. Mod. Phys.}\ }\textbf {\bibinfo
  {volume} {D23}},\ \bibinfo {pages} {1430001} (\bibinfo {year} {2014})},\
  \Eprint {http://arxiv.org/abs/1309.2557} {arXiv:1309.2557 [hep-th]}
  \BibitemShut {NoStop}%
\bibitem [{\citenamefont {Penedones}(2011)}]{Penedones:2010ue}%
  \BibitemOpen
  \bibfield  {author} {\bibinfo {author} {\bibfnamefont {J.}~\bibnamefont
  {Penedones}},\ }\href {\doibase 10.1007/JHEP03(2011)025} {\bibfield
  {journal} {\bibinfo  {journal} {JHEP}\ }\textbf {\bibinfo {volume} {03}},\
  \bibinfo {pages} {025} (\bibinfo {year} {2011})},\ \Eprint
  {http://arxiv.org/abs/1011.1485} {arXiv:1011.1485 [hep-th]} \BibitemShut
  {NoStop}%
\bibitem [{\citenamefont {Fitzpatrick}\ and\ \citenamefont
  {Kaplan}(2012)}]{Fitzpatrick:2011hu}%
  \BibitemOpen
  \bibfield  {author} {\bibinfo {author} {\bibfnamefont {A.~L.}\ \bibnamefont
  {Fitzpatrick}}\ and\ \bibinfo {author} {\bibfnamefont {J.}~\bibnamefont
  {Kaplan}},\ }\href {\doibase 10.1007/JHEP10(2012)127} {\bibfield  {journal}
  {\bibinfo  {journal} {JHEP}\ }\textbf {\bibinfo {volume} {10}},\ \bibinfo
  {pages} {127} (\bibinfo {year} {2012})},\ \Eprint
  {http://arxiv.org/abs/1111.6972} {arXiv:1111.6972 [hep-th]} \BibitemShut
  {NoStop}%
\bibitem [{\citenamefont {Aharony}\ \emph {et~al.}(2017)\citenamefont
  {Aharony}, \citenamefont {Alday}, \citenamefont {Bissi},\ and\ \citenamefont
  {Perlmutter}}]{Aharony:2016dwx}%
  \BibitemOpen
  \bibfield  {author} {\bibinfo {author} {\bibfnamefont {O.}~\bibnamefont
  {Aharony}}, \bibinfo {author} {\bibfnamefont {L.~F.}\ \bibnamefont {Alday}},
  \bibinfo {author} {\bibfnamefont {A.}~\bibnamefont {Bissi}}, \ and\ \bibinfo
  {author} {\bibfnamefont {E.}~\bibnamefont {Perlmutter}},\ }\href {\doibase
  10.1007/JHEP07(2017)036} {\bibfield  {journal} {\bibinfo  {journal} {JHEP}\
  }\textbf {\bibinfo {volume} {07}},\ \bibinfo {pages} {036} (\bibinfo {year}
  {2017})},\ \Eprint {http://arxiv.org/abs/1612.03891} {arXiv:1612.03891
  [hep-th]} \BibitemShut {NoStop}%
\bibitem [{\citenamefont {Rastelli}\ and\ \citenamefont
  {Zhou}(2017)}]{Rastelli:2016nze}%
  \BibitemOpen
  \bibfield  {author} {\bibinfo {author} {\bibfnamefont {L.}~\bibnamefont
  {Rastelli}}\ and\ \bibinfo {author} {\bibfnamefont {X.}~\bibnamefont
  {Zhou}},\ }\href {\doibase 10.1103/PhysRevLett.118.091602} {\bibfield
  {journal} {\bibinfo  {journal} {Phys. Rev. Lett.}\ }\textbf {\bibinfo
  {volume} {118}},\ \bibinfo {pages} {091602} (\bibinfo {year} {2017})},\
  \Eprint {http://arxiv.org/abs/1608.06624} {arXiv:1608.06624 [hep-th]}
  \BibitemShut {NoStop}%
\bibitem [{\citenamefont {Cardona}(2017)}]{Cardona:2017tsw}%
  \BibitemOpen
  \bibfield  {author} {\bibinfo {author} {\bibfnamefont {C.}~\bibnamefont
  {Cardona}},\ }\href@noop {} {\  (\bibinfo {year} {2017})},\ \Eprint
  {http://arxiv.org/abs/1708.06339} {arXiv:1708.06339 [hep-th]} \BibitemShut
  {NoStop}%
\bibitem [{\citenamefont {Giombi}\ \emph {et~al.}(2018)\citenamefont {Giombi},
  \citenamefont {Sleight},\ and\ \citenamefont {Taronna}}]{Giombi:2017hpr}%
  \BibitemOpen
  \bibfield  {author} {\bibinfo {author} {\bibfnamefont {S.}~\bibnamefont
  {Giombi}}, \bibinfo {author} {\bibfnamefont {C.}~\bibnamefont {Sleight}}, \
  and\ \bibinfo {author} {\bibfnamefont {M.}~\bibnamefont {Taronna}},\ }\href
  {\doibase 10.1007/JHEP06(2018)030} {\bibfield  {journal} {\bibinfo  {journal}
  {JHEP}\ }\textbf {\bibinfo {volume} {06}},\ \bibinfo {pages} {030} (\bibinfo
  {year} {2018})},\ \Eprint {http://arxiv.org/abs/1708.08404} {arXiv:1708.08404
  [hep-th]} \BibitemShut {NoStop}%
\bibitem [{\citenamefont {Yuan}(2018)}]{Yuan:2018qva}%
  \BibitemOpen
  \bibfield  {author} {\bibinfo {author} {\bibfnamefont {E.~Y.}\ \bibnamefont
  {Yuan}},\ }\href@noop {} {\  (\bibinfo {year} {2018})},\ \Eprint
  {http://arxiv.org/abs/1801.07283} {arXiv:1801.07283 [hep-th]} \BibitemShut
  {NoStop}%
\bibitem [{Note1()}]{Note1}%
  \BibitemOpen
  \bibinfo {note} {\begin {small}The cubic coupling is known to be extremal,
  and the simplest $\Yup $-diagram is divergent\end {small}}\BibitemShut
  {NoStop}%
\bibitem [{\citenamefont {Beisert}\ \emph {et~al.}(2012)\citenamefont {Beisert}
  \emph {et~al.}}]{Beisert:2010jr}%
  \BibitemOpen
  \bibfield  {author} {\bibinfo {author} {\bibfnamefont {N.}~\bibnamefont
  {Beisert}} \emph {et~al.},\ }\href@noop {} {\bibfield  {journal} {\bibinfo
  {journal} {Lett. Math. Phys.}\ }\textbf {\bibinfo {volume} {99, 3}} (\bibinfo
  {year} {2012})},\ \Eprint {http://arxiv.org/abs/1012.3982} {arXiv:1012.3982
  [hep-th]} \BibitemShut {NoStop}%
\bibitem [{Note2()}]{Note2}%
  \BibitemOpen
  \bibinfo {note} {\begin {small}$\protect \mathrm {d}\mu _{x_1,x_2,\protect
  \ldots }=\protect \mathrm {d}^4 x_1 \protect \sqrt {|g(x_1)|}\times \protect
  \mathrm {d}^4x_2 \protect \sqrt {|g(x_2)|}\protect \dots $\end
  {small}}\BibitemShut {NoStop}%
\bibitem [{\citenamefont {Witten}(1998)}]{Witten:1998qj}%
  \BibitemOpen
  \bibfield  {author} {\bibinfo {author} {\bibfnamefont {E.}~\bibnamefont
  {Witten}},\ }\href@noop {} {\bibfield  {journal} {\bibinfo  {journal} {Adv.
  Theor. Math. Phys.}\ }\textbf {\bibinfo {volume} {2}},\ \bibinfo {pages}
  {253} (\bibinfo {year} {1998})},\ \Eprint
  {http://arxiv.org/abs/hep-th/9802150} {arXiv:hep-th/9802150} \BibitemShut
  {NoStop}%
\bibitem [{\citenamefont {Freedman}\ \emph {et~al.}(1999)\citenamefont
  {Freedman}, \citenamefont {Mathur}, \citenamefont {Matusis},\ and\
  \citenamefont {Rastelli}}]{Freedman:1998tz}%
  \BibitemOpen
  \bibfield  {author} {\bibinfo {author} {\bibfnamefont {D.~Z.}\ \bibnamefont
  {Freedman}}, \bibinfo {author} {\bibfnamefont {S.~D.}\ \bibnamefont
  {Mathur}}, \bibinfo {author} {\bibfnamefont {A.}~\bibnamefont {Matusis}}, \
  and\ \bibinfo {author} {\bibfnamefont {L.}~\bibnamefont {Rastelli}},\ }\href
  {\doibase 10.1016/S0550-3213(99)00053-X} {\bibfield  {journal} {\bibinfo
  {journal} {Nucl. Phys.}\ }\textbf {\bibinfo {volume} {B546}},\ \bibinfo
  {pages} {96} (\bibinfo {year} {1999})},\ \Eprint
  {http://arxiv.org/abs/hep-th/9804058} {arXiv:hep-th/9804058} \BibitemShut
  {NoStop}%
\bibitem [{\citenamefont {Mueck}\ and\ \citenamefont
  {Viswanathan}(1998)}]{Muck:1998rr}%
  \BibitemOpen
  \bibfield  {author} {\bibinfo {author} {\bibfnamefont {W.}~\bibnamefont
  {Mueck}}\ and\ \bibinfo {author} {\bibfnamefont {K.~S.}\ \bibnamefont
  {Viswanathan}},\ }\href {\doibase 10.1103/PhysRevD.58.041901} {\bibfield
  {journal} {\bibinfo  {journal} {Phys. Rev.}\ }\textbf {\bibinfo {volume}
  {D58}},\ \bibinfo {pages} {041901} (\bibinfo {year} {1998})},\ \Eprint
  {http://arxiv.org/abs/hep-th/9804035} {arXiv:hep-th/9804035} \BibitemShut
  {NoStop}%
\bibitem [{\citenamefont {Maldacena}(1998)}]{Maldacena:1997re}%
  \BibitemOpen
  \bibfield  {author} {\bibinfo {author} {\bibfnamefont {J.~M.}\ \bibnamefont
  {Maldacena}},\ }\href@noop {} {\bibfield  {journal} {\bibinfo  {journal}
  {Adv. Theor. Math. Phys.}\ }\textbf {\bibinfo {volume} {2}},\ \bibinfo
  {pages} {231} (\bibinfo {year} {1998})},\ \Eprint
  {http://arxiv.org/abs/hep-th/9711200} {arXiv:hep-th/9711200} \BibitemShut
  {NoStop}%
\bibitem [{\citenamefont {Gubser}\ \emph {et~al.}(1998)\citenamefont {Gubser},
  \citenamefont {Klebanov},\ and\ \citenamefont {Polyakov}}]{Gubser:1998bc}%
  \BibitemOpen
  \bibfield  {author} {\bibinfo {author} {\bibfnamefont {S.~S.}\ \bibnamefont
  {Gubser}}, \bibinfo {author} {\bibfnamefont {I.~R.}\ \bibnamefont
  {Klebanov}}, \ and\ \bibinfo {author} {\bibfnamefont {A.~M.}\ \bibnamefont
  {Polyakov}},\ }\href {\doibase 10.1016/S0370-2693(98)00377-3} {\bibfield
  {journal} {\bibinfo  {journal} {Phys. Lett.}\ }\textbf {\bibinfo {volume}
  {B428}},\ \bibinfo {pages} {105} (\bibinfo {year} {1998})},\ \Eprint
  {http://arxiv.org/abs/hep-th/9802109} {arXiv:hep-th/9802109} \BibitemShut
  {NoStop}%
\bibitem [{\citenamefont {Dolan}\ and\ \citenamefont
  {Osborn}(2001)}]{Dolan:2000ut}%
  \BibitemOpen
  \bibfield  {author} {\bibinfo {author} {\bibfnamefont {F.~A.}\ \bibnamefont
  {Dolan}}\ and\ \bibinfo {author} {\bibfnamefont {H.}~\bibnamefont {Osborn}},\
  }\href {\doibase 10.1016/S0550-3213(01)00013-X} {\bibfield  {journal}
  {\bibinfo  {journal} {Nucl. Phys.}\ }\textbf {\bibinfo {volume} {B599}},\
  \bibinfo {pages} {459} (\bibinfo {year} {2001})},\ \Eprint
  {http://arxiv.org/abs/hep-th/0011040} {arXiv:hep-th/0011040} \BibitemShut
  {NoStop}%
\end{thebibliography}%
\end{document}